\documentclass[journal=jacsat,manuscript=article]{achemso}

\usepackage[version=3]{mhchem} 
\usepackage[utf8]{inputenc}
\usepackage{graphicx}
\usepackage{placeins}
\usepackage{float}
\usepackage{subfigure}
\usepackage{color,soul}
\usepackage{mathtools}
\usepackage{amsmath}

\newcommand{\DC}{$^\circ$C}
\author{Guy Jacoby}
\affiliation{Raymond \& Beverly Sackler School of Physics \& Astronomy, Tel Aviv University, Tel Aviv 6997801, Israel}
\alsoaffiliation[biosoft]{The Center for Physics and Chemistry of Living Systems, Tel Aviv University, Israel}
\alsoaffiliation[nano]{The Center for NanoTechnology and NanoScience, Tel Aviv Univeristy, Israel}
\author{Merav Segal Asher}
\affiliation{Raymond \& Beverly Sackler School of Chemistry, Tel Aviv University, Tel Aviv 6997801, Israel}
\alsoaffiliation[biosoft]{The Center for Physics and Chemistry of Living Systems, Tel Aviv University, Israel}
\alsoaffiliation[nano]{The Center for NanoTechnology and NanoScience, Tel Aviv Univeristy, Israel}
\author{Tamara Ehm}
\affiliation{Raymond \& Beverly Sackler School of Physics \& Astronomy, Tel Aviv University, Tel Aviv 6997801, Israel}
\alsoaffiliation[LMU]{ Faculty of Physics and Center for NanoScience, Ludwig-Maximilians-Universität, München, Germany}
\alsoaffiliation[biosoft]{The Center for Physics and Chemistry of Living Systems, Tel Aviv University, Israel}
\alsoaffiliation[nano]{The Center for NanoTechnology and NanoScience, Tel Aviv Univeristy, Israel}
\author{Inbal Abutbul Ionita}
\affiliation[technion]{CryoEM Laboratory of Soft Matter, Faculty of Biotechnology and Food Engineering, Technion-Israel Institute of Technology, Haifa 3200003, Israel}
\author{Hila Shinar}
\affiliation{Raymond \& Beverly Sackler School of Physics \& Astronomy, Tel Aviv University, Tel Aviv 6997801, Israel}
\alsoaffiliation[biosoft]{The Center for Physics and Chemistry of Living Systems, Tel Aviv University, Israel}
\alsoaffiliation[nano]{The Center for NanoTechnology and NanoScience, Tel Aviv Univeristy, Israel}
\author{Salome Azoulay-Ginsburg}
\affiliation{Raymond \& Beverly Sackler School of Chemistry, Tel Aviv University, Tel Aviv 6997801, Israel}
\author{Dganit Danino}
\affiliation[technion]{CryoEM Laboratory of Soft Matter, Faculty of Biotechnology and Food Engineering, Technion-Israel Institute of Technology, Haifa 3200003, Israel}
\alsoaffiliation[china]{Guangdong-Technion Israel Institute of Technology, Shantou, Guangdong Province, 515063 China}
\author{Michael M. Kozlov}
\affiliation{Sackler School of Medicine, Tel Aviv University, Tel Aviv 6997801, Israel}
\alsoaffiliation[biosoft]{The Center for Physics and Chemistry of Living Systems, Tel Aviv University, Israel}

\author{Roey J. Amir}
\affiliation{Raymond \& Beverly Sackler School of Chemistry, Tel Aviv University, Tel Aviv 6997801, Israel}
\alsoaffiliation[biosoft]{The Center for Physics and Chemistry of Living Systems, Tel Aviv University, Israel}
\alsoaffiliation[nano]{The Center for NanoTechnology and NanoScience, Tel Aviv Univeristy, Israel}
\email{amirroey@tauex.tau.ac.il}
\author{Roy Beck}
\affiliation{Raymond \& Beverly Sackler School of Physics \& Astronomy, Tel Aviv University, Tel Aviv 6997801, Israel}
\alsoaffiliation[biosoft]{The Center for Physics and Chemistry of Living Systems, Tel Aviv University, Israel}
\alsoaffiliation[nano]{The Center for NanoTechnology and NanoScience, Tel Aviv Univeristy, Israel}
\email{roy@tauex.tau.ac.il}

\title[An \textsf{achemso} demo]
{Order from disorder with intrinsically disordered peptide amphiphiles}

\abbreviations{IR,NMR,UV}

\begin{document}

\begin{tocentry}

\begin{figure}[H]
\includegraphics[width=0.6\textwidth]{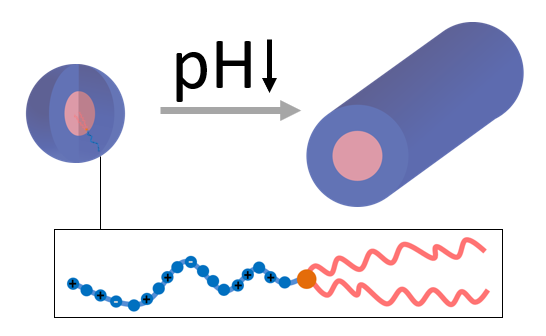}

\end{figure}

\end{tocentry}

\begin{abstract}
Amphiphilic molecules and their self-assembled structures have long been the target of extensive research due to their potential applications in fields ranging from materials design to biomedical and cosmetic applications. Increasing demands for functional complexity have been met with challenges in biochemical engineering, driving researchers to innovate in the design of new amphiphiles. An emerging class of molecules, namely, peptide amphiphiles, combines key advantages and circumvents some of the disadvantages of conventional phospholipids and block-copolymers. Herein, we present new peptide amphiphiles comprised of an intrinsically disordered peptide conjugated to two variants of hydrophobic dendritic domains. These molecules termed intrinsically disordered peptide amphiphiles (IDPA), exhibit a sharp pH-induced micellar phase-transition from low-dispersity spheres to extremely elongated worm-like micelles. We present an experimental characterization of the transition and propose a theoretical model to describe the pH-response. We also present the potential of the shape transition to serve as a mechanism for the design of a cargo hold-and-release application. Such amphiphilic systems demonstrate the power of tailoring the interactions between disordered peptides for various stimuli-responsive biomedical applications.  
\end{abstract}

\section{Introduction}

Self-assembly of amphiphilic molecules holds great interest from a fundamental scientific point of view, as well as for their potential for creating nanocarriers for various applications, ranging from drugs and nucleic acids in medicine to fragrances and other small chemicals in the food and cosmetics industries \cite{Kulkarni2012}. The functionality of nanocarriers' design will depend on many factors, including the mechanism for cargo hold-and-release, bio-compatibility, uniformity, and tunability, all of which present challenging obstacles in engineering efficient nanocarriers. Natural lipids and synthetic block-copolymers are two of the most widely used amphiphiles in designing such systems, each prevailing due to its specific advantages \cite{dengler2013mesoporous,Rodriguez2005, gunatillake2003biodegradable,segal2020architectural,harnoy2014enzyme,peer2007,Mai2012,Carlsen2009,Zhuang2013,Miyata2011,Segal2017}

An emerging class of synthetic amphiphiles, namely, peptide amphiphiles (PA), is designed to self-assemble into functional structures by building upon the advantageous characteristics of lipids and block-copolymers \cite{uversky2009mysterious,he2009predicting,uversky2017intrinsically}. The molecules' hydrophilic domain, usually a bio-inspired peptide engineered to fulfill one or more active roles, is chemically conjugated to a hydrophobic tail group, usually single or double chain fatty acids like those found in lipids. Due to their highly flexible design scheme, PA self-assemblies can act as organic scaffolds for bone-like mineralization \cite{Hartgerink2001}, anisotropic actuators mimicking skeletal muscle \cite{Chin2018}, produce new versatile soft materials \cite{Hartgerink2002}, and enhance neural progenitor cell differentiation into neurons \cite{Silva2004}. PA can contain spacers or linkers in their design, connecting the hydrophobic tails to the functional hydrophilic domain \cite{Peters2009, Acar2017}, or conjugating to an "external" functional group, such as an MRI contrast agent \cite{Bull2005}. Comprehensive research is done to understand the properties and roles of the different molecular domains and their contribution to the specific self-assembled mesophase \cite{Tovar2005,Paramonov2006}. In many cases, the bio-inspired peptides are designed or derived from proteins with some degree of secondary structure \cite{Trent2011}. Nonetheless, PA hydrophilic domains are not exclusively bio-inspired and can be made of any polypeptide sequence, including intrinsically disordered (i.e., unfolded) peptide sequences.

An increasingly large number of proteins have been found to lack a fixed or ordered structure \cite{Oates2013}. As such, these proteins have been termed intrinsically disordered proteins (IDPs). A more operational definition of an IDP is a protein that does not possess only a single functional conformation, but rather it can fold into an ensemble of functional conformations depending on the setting. This structural plasticity is, in many cases, the result of relatively weaker and transient interactions between proteins segments \cite{kornreich2015c,Kornreich2016,Malka-Gibor2017a,DeForte2016}. The nature of their interactions makes IDPs better suited for specific roles than structured proteins as they can interact with multiple partners or retain in liquid condensed phase \cite{Wei2017,Shin2017}.

Envisioning systems with intricate and precise self-organization potential, as sought after in applications, requires the fabrication and processing of unique nanostructures. A primary limitation of current block-copolymer synthesis is the lack of general methods for producing precise chain structure (i.e., sequence control) to facilitate multiple desired functions. Natural lipids also have limited functionality due to their relatively small hydrophilic domain. In contrast, intrinsically disordered peptide amphiphiles (IDPAs) benefit from the use of relatively short sequences that are easier to synthesize while still retaining rich functionality. PAs can combine the functionality and flexibility of peptides; since there are 20 natural amino acids, there are practically numerous ($>10^{9}$) possible sequences, even for short (18mers) peptides such as the one studied in this work. Naturally, the design of IDP based hydrophilic domains can be inspired by biology, utilizing the immense pre-existing knowledge base of proteins.  
To date, only few examples demonstrated exciting functionality using disordered domain in PA \cite{Chin2018,Dzuricky2020,Silva2004,Peters2009,mozhdehi2018genetically}.

\begin{figure*}
\includegraphics[width=1\textwidth]{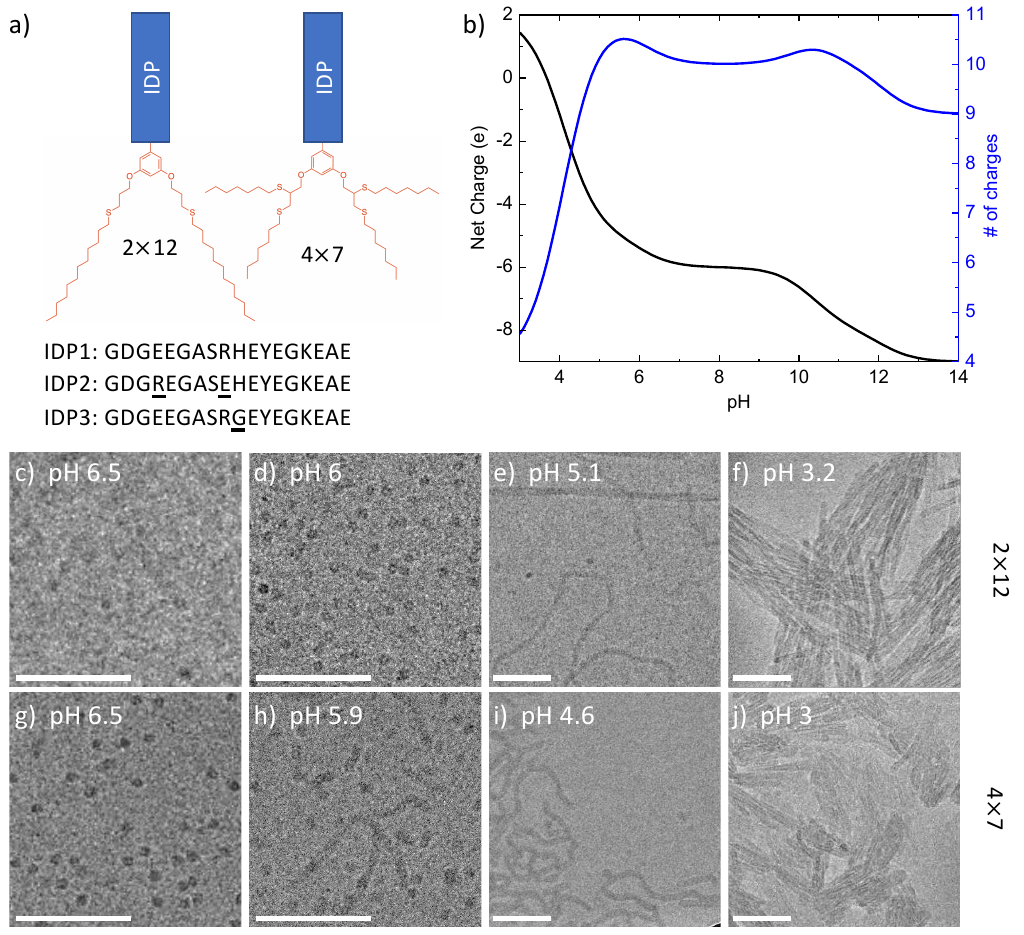}
\caption{(a) Schematics of the IDPAs with two tail variants (2$\times$12  and 4$\times$7) and the one-letter IDP's sequences used in this study. (b) Net charge (black) and the number of charged amino acids (blue) of \textit{IDPA1} hydrophilic domain as a function of pH. (c-j) \textit{IDPA1} Cryo-TEM images showing of self-assembly of (c-f) 2$\times$12 and (g-j) 4$\times$7 at various marked pHs. Cryo-TEM images show (c,d,g) spherical micelles at low pH, (e,h,i) coexistence with worm-like micelles at intermediate pH, and (f,j) aggregated micellar rods at low pH. Scale bar is 100 nm}
  \label{Intro}
\end{figure*}
Here, we study the self-assembly and encapsulation capabilities of two IDPAs, composed of hydrophilic domain inspired by the neurofilament-low disordered tail domain and a dendritic branching unit used for conjugating the lipophilic tails\cite{Uversky2010,Kornreich2016,Kornreich2015}. Using turbidity, Small-angle X-ray Scattering (SAXS) and cryogenic transmission electron microscopy (cryo-TEM) measurements, we found that the IDPAs self-assembled into well-defined, low-dispersity nanoparticles. In addition, we show that the IDPAs' sensitivity to pH leads to a tunable and robust organization of the self-assembled nanoparticles. We further show that minor alterations in the peptide sequence can lead to alteration in IDPA-IDPA interaction and the macroscopic arrangement. Last, we demonstrate the potential of using the pH induced shape transition as a release mechanism for the design of nano carriers. 

\section{Results and discussion}

\subsection{Synthesis and the structure of IDPAs}
IDP sequences were synthesized on an automated solid-phase peptide synthesizer using Fmoc-protected amino acids. The hydrophilic peptide domain sequence is an 18mer amino-acid polyampholyte, inspired by the intrinsically disordered carboxy domain of the protein neurofilament-low (NF-L) \cite{Kornreich2015,Kornreich2016,Malka-Gibor2017a,Laser-Azogui2015}. Once the IDP sequences synthesis was completed, an aromatic branching unit containing two allyl or propargyl functionalities and a carboxylic acid was used to cap the N-terminus of the IDP sequence. After the branching units were conjugated, the capped peptides were cleaved from the resin using TFA, and hydrophobic end-groups containing thiols (dodecane-thiol and heptane-thiol) were conjugated to the allyl or propargyl moieties through thiol-ene or -yne click reactions, respectively.\cite{hoyle2010thiol,kade2010power,konkolewicz2009hyperbranched,lowe2014thiol} We term the tail-group variants by 2$\times$12 and 4$\times$7, to represent the number and length of the alkyl chains  and the amphiphiles as \textit{IDPA1} (Fig. \ref{Intro}a). The one letter amino acid sequence of \textit{IDPA1} is GDGEEGASRHEYEGKEAE. 

Notably, the peptide's sequence includes 11 protonatable residues, allowing for the net charge of the peptide to vary significantly as a function of pH (Fig. \ref{Intro}b). Specifically, at approximately pH 5.5 there is a decrease in the net charge and in the number of charged residues due to Aspartic Acid and Glutamic Acid residues' protonation. The critical micelle concentrations (CMC) of 11 $\mu$M were determined using the solvatochromic dye Nile Red (Supplementary Figure S1). A CMC of 5 $\mu$M is in the typical micro-molar range for peptide amphiphiles \cite{Acar2017,Vincenzi2015,Klass2019,Black2012}.

The peptide's degree of disorder was experimentally verified by measuring the circular dichroism (CD) spectrum of samples of the peptide (unconjugated) and the two \textit{IDPA1} variants (Supplementary Figure S2). In addition, the peptide sequence displays a high probability for disorder and the absence of regular secondary structure using the NetSurfP-2.0 bioinformatic algorithm (Supplementary Figure S3) \cite{Klausen2019}.

\subsection{Micellar nanostructures at high pH}

We expected \textit{IDPA1} to show little to no sensitivity at high pH where the peptides' net charge state remains constant (Fig. \ref{Intro}b). Indeed, above pH$\sim$6, we find that both tail-variants assemble into nanoscopic spherical micelles, which were visualized via cryo-TEM (Fig. \ref{Intro}c-j). At this slightly acidic pH, the micelles showed repulsion and remained miscible at a relatively high IDPA concentration (10 mg/ml). SAXS revealed low-dispersity spherical nanostructures, typical for structured particles. A core-shell form-factor was used to fit the SAXS data and provided the radius of the hydrophobic core $R_{core}=1.25\pm 0.09$ nm, the width of the peptide shell (hydrophilic domain region surrounding the core) $w_{shell}=2.12\pm 0.05$ nm, and the respective average electron densities, $\rho_{core}=284$ $e\slash nm^{3}$ and $\rho_{shell}=355$ $e\slash nm^{3}$ (Fig. \ref{SAXSdata}b). Using the SAXS fit and the values from Harpaz et al. \cite{Harpaz1994}, we estimate the number of monomers for 4$\times$7 and 2$\times$12 \textit{IDPA1} at pH 7.5 to be approximately 13 and 40, respectively.

The SAXS measurements further revealed the composite dimensionality of the spherical micelles. Using the Kratky analysis \cite{Orthaber2000}, we found a bell-shaped curve at lower $q$, which corresponds to the 3D nature of the micelle at larger length-scales. However, the linear increase at larger $q$ suggests an unfolded state of the peptides at smaller length-scales (Fig. \ref{SAXSdata}c). Approaching pH$\sim$6, we find a mild alteration in the SAXS signal, indicating a structural rearrangement at lower pH that we discuss next.  

\subsection{pH induced phase-transition}
\begin{figure*}
\includegraphics[width=1\textwidth]{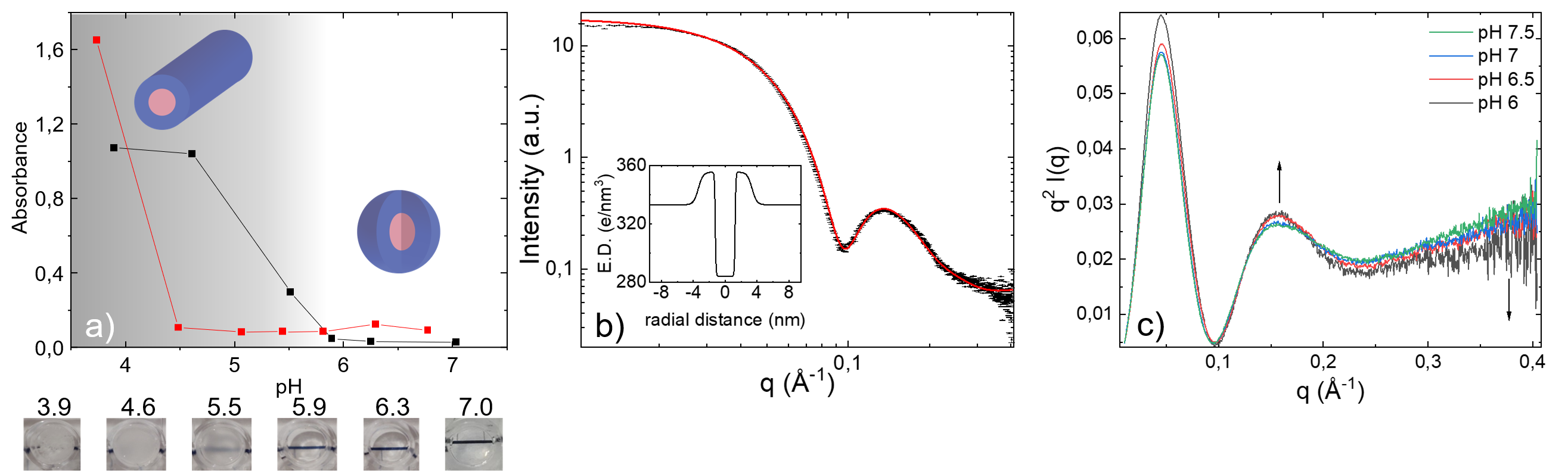}
\caption{(a) Turbidity measurement of 4$\times$7 (black) and 2$\times$12 (red) \textit{IDPA1} with schematic representation of phase transition from spherical to cylindrical micelles. Results show an increase in turbidity when lowering the pH, indicating a transition into large assemblies. Below: photographs of the 4$\times$7 \textit{IDPA1} samples measured in the experiment (numbers above photos indicate the pH). b) Spherical core-shell form-factor fit for the SAXS data. Inset, electron density profile used in the fit. c)  Kratky analysis with a bell-shaped curve at lower $q$, corresponding to 3D nature of the micelle at larger length-scales, and linear increase at larger $q$ resulted from the unfolded state of the peptides at smaller length-scales.} 
  \label{SAXSdata}
\end{figure*}

The peptide sequence is a strong polyampholyte; hence, the self-assembly of \textit{IDPA1} is expected to depend on pH. Therefore, lowering the pH towards the pI can facilitate a structural phase-transition that depends on the IDP charge density and the amphiphiles' electrostatic interaction. Turbidity measurements performed on both IDPAs revealed a clear difference in sample translucency above and below pH 6 (Fig. \ref{SAXSdata}a). This transition indicates a macro-molecular aggregation of the self-assembled structures at low pH. Below pH 6, the nanostructures interact to produce larger and more ordered aggregates. Figure \ref{SAXSdata}a showcases the transition from translucent to opaque solutions. The turbidity of both variants changes abruptly at pH$\sim$ 6 and 4.5, for the 4$\times$7 and 2$\times$12, respectively.   

Cryo-TEM micrographs of \textit{IDPA1} variants in the pH range 3-6.5 confirm the microscopic transition hinted at by the turbidity measurements (Fig. \ref{Intro}c-j). The micrographs show a transition from spherical micelles at pH 6.5 to elongated worm-like micelles lower than pH 5.5. Furthermore, at lower pHs the worm-like micelles are strongly interacting and aggregating. In the case of 4$\times$7, the image taken at pH 5.9 nicely shows the coexistence of spherical and worm-like micelles (Fig. \ref{Intro}h). In addition, the sphere-to-rod transition and micellar growth is earlier for the 4$\times$7 variant and is consistent with its higher hydrophobicity and turbidity measurements. 

Further verification of the phase transition is clearly shown using SAXS (Figs. \ref{SAXS}a,b). At low pH, the scattering is qualitatively different and is no longer a sum of independent spherical scatterers producing a form-factor SAXS signal. Instead, the SAXS signal now includes an additional structure-factor signal produced by the inter-particles' correlation. The structure-factor peaks position match a 2D hexagonal lattice for 2$\times$12 and a 1D lattice for 4$\times$7 with the corresponding unit-cell spacing of $d_H=10.2$ nm and $d_L=9$ nm. The highly-dense packing is also evident bythe cryo-TEM micrographs showing organization of the micellar rods at low pHs (Fig. \ref{Intro}f,j).

\subsection{Engineering the self-assembly by point mutation}

The added value of using peptides as the hydrophilic domain is the possibility to tune the interactions via small alterations in the sequence, such as a single mutation. We recently demonstrated that a similar sequence peptide alters its self-interaction via a single point mutation \cite{Chakraborty2019}. Short-ranged transient interactions are also present in the original neurofilament disordered protein \cite{Kornreich2016, Malka-Gibor2017a, Kornreich2015a, Pregent2015, Morgan2020, Beck2010}, presumably due to the oppositely charged amino-acids along the polyampholytic intrinsically disordered C-terminus domain. We synthesized a sequence variant by changing the Glutamic Acid at position 4 with the Arginine at position 9, termed \textit{IDPA2}. By doing so, we are conserving the net charge and the functional relation between pH and charge but altering the charge distribution at the hydrophilic peptide domain. Such alteration is expected to alter the ionic bridging between the IDPs \cite{Beck2010,Kornreich2016,Kornreich2015a}. 

\begin{figure*}
\includegraphics[width=1\textwidth]{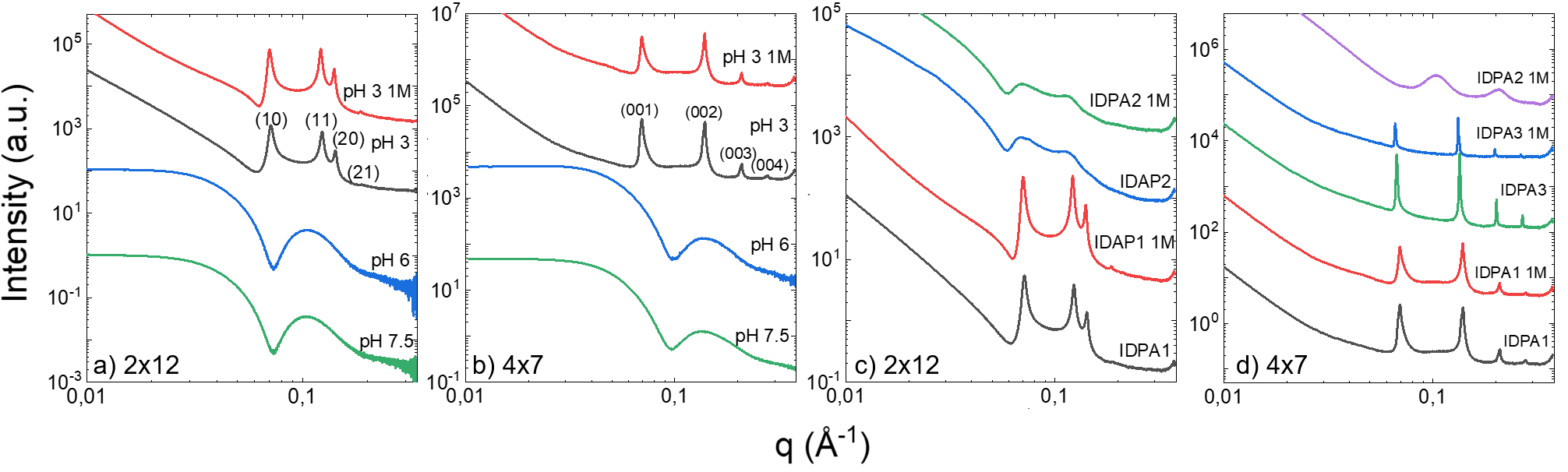}
\caption{SAXS measurements of (a) 2$\times$12 and (b) 4$\times$7 \textit{IDPA1} variants at different pHs. Above pH 6 the scattering profile pertains to spherical micelles. Below pH 6, the scattering is dominated by a structure-factor. The 2$\times$12 variant forms a hexagonal phase with a spacing of $d_H=10.2$, stable at  1M NaCl (red line). For the 4$\times$7 variant, the scattering at low pHs is dominated by a 1D phase structure with a spacing of $d_L=9$ nm.(c) Comparison of SAXS measurements of 2$\times$12 for \textit{IDPA1} and \textit{IDPA2} at pH 3, in either 150 mM salt or 1 M (labeled). The small change in charge distribution has a dramatic effect on the interaction of the worm-like micelles. The sharp structure-factor peaks are replaced with wide and shallow peaks, indicating weaker correlations. (d) Comparison of SAXS measurements of 4$\times$7 variants at pH 3, in either 150 mM salt or 1 M (labeled). The mesophase remains the same for \textit{IDPA1} and \textit{IDPA3}, while the \textit{IDPA2} variant shows a pronounced weakening of the inter-micelle correlations.}
  \label{SAXS}
\end{figure*}

In addition, when considering the observed structural pH sensitivity, with a phase-transition being located somewhere between pH 4 and 6 (Figs. \ref{Intro}c,d and \ref{SAXS}), a possible origin for the transition can be the charging state of the Histidine amino-acid. To verify this hypothesis, we designed a slightly modified hydrophilic domain sequence, termed \textit{IDPA3}, that replaces the Histidine at position 10 with a neutral and pH insensitive Glycine residue. Notably, the two new IDPA variants are still considered highly disordered and are assigned the random coil conformation by the predictors (Supplementary Figure S4).

Indeed, we found that the hydrophilic domain (i.e., the disordered peptide) and its interactions in the assembly control the complex aggregations at low pH. SAXS experiments show that replacing the Histidine with a Glycine served to strengthen the interaction between the worm-like micelles (Fig. \ref{SAXS}d). However, slightly changing the sequence's order, namely the \textit{IDPA2} variant, noticeably changes the complex aggregation (Fig. \ref{SAXS}c). For the 2$\times$12 variant, we saw a significant weakening of micelle-micelle interactions in the worm-like phase, at 150 mM and 1 M salt concentration, demonstrated by the diminished structure-factor scattering (Fig. \ref{SAXS}a). However, the 4$\times$7 variant showed a similar weakening of the interactions only at 1 M salt pointing towards non-trivial electrostatic interaction \cite{Kornreich2016,Beck2010,kornreich2015c} between the IDPs (Fig. \ref{SAXS}b).

\subsection{Cargo encapsulation and release}
\begin{figure}
\includegraphics[width=0.8\textwidth]{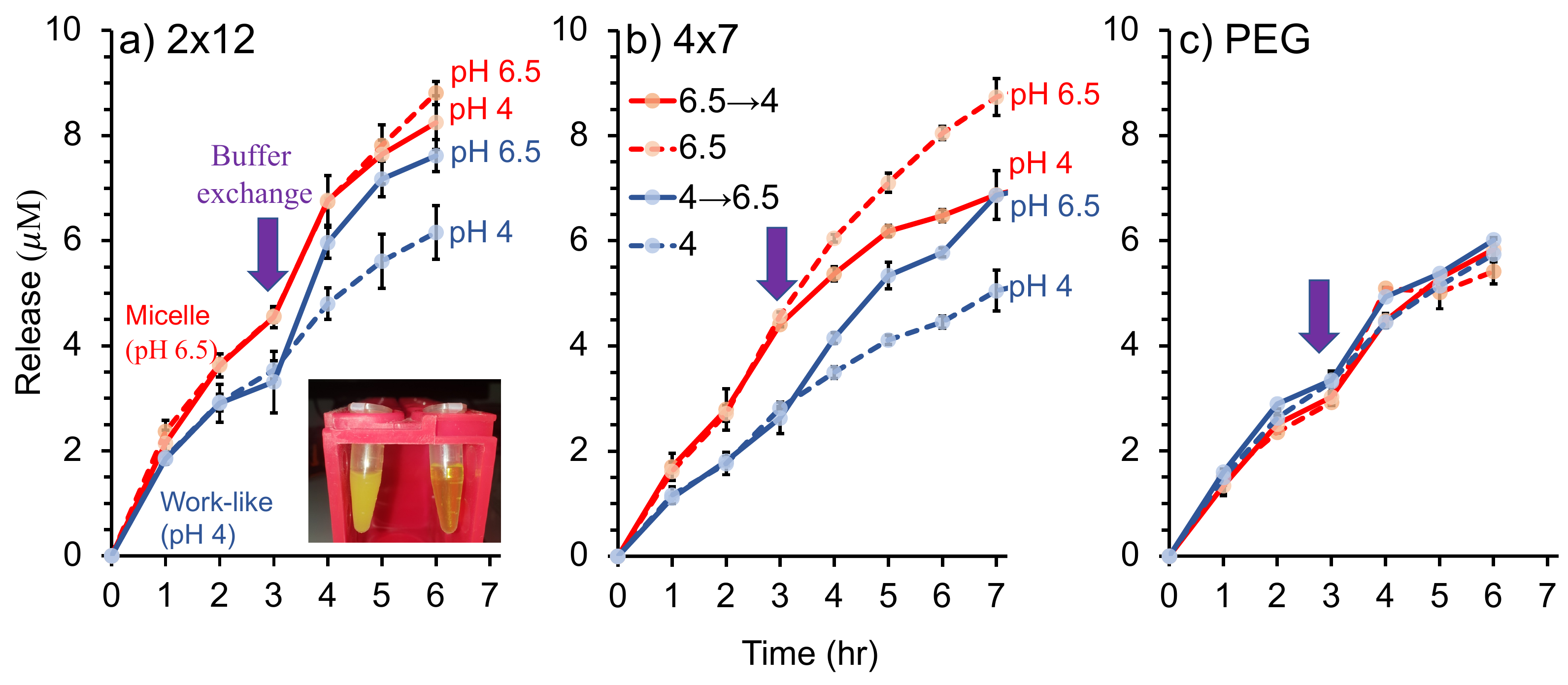}
 \caption{Encapsulation measurements. Accumulative released dye concentration for (a) 2$\times$12, (b) 4$\times$7 IDPAs and (c) PEG-2$\times$12. Blue and red data-points represent a baseline pH for the experiments of 6.5 and 4, respectively. After 3 hours the buffer was exchanged either to induce structural mesophase transition via pH trigger (solid lines), or to identical and fresh buffer (dashed line). PEG-2$\times$12 amphiphiles at pH 6.5 and pH 4 show no pH release trigger. Inset, representative photo of the encapsulated dye in (left) worm-like micelles at pH 4 and (right) spherical micelles at pH 6.5.}
  \label{encapsulation}
\end{figure}

After the pH-depended self-assembly into micelles and worm-like micelles was characterized, we wanted to evaluate how the shape transition will affect encapsulated cargo release. We chose butyl ester of 7-(Diethylamino)coumarin-3-carboxylic acid as the hydrophobic cargo and used a dialysis set-up to study the hydrophobic dye's release from the assembled structures. Solutions containing micelles of 2$\times$12 and 4$\times$7 IDPAs (pH 6.5) were mixed with a stock solution of the dye, followed by filtration of the residual un-encapsulated dyes. Next, the samples' pH was adjusted to pH 4 by adding a few $\mu$L of HCl to transform the spherical micellar assemblies into the worm-like micelles. As expected, while the micelle solution was clear, the worm-like micelle solution became highly turbid (Fig. \ref{encapsulation}). 

Next, the two solutions were transferred into dialysis tubes and placed in buffer solutions (pH 6.5 or 4 for the micelles and worm-like micelles, respectively) containing bovine serum albumin (BSA). The BSA's role was to scavenge the released dyes and avoid their aggregation and dissolution due to their poor aqueous solubility. The solutions were then placed in a shaking incubator at 37$^{\circ}$C, and samples were taken periodically from the outer solution and characterized by a spectrophotometer to determine the released dye's concentration.  

We find that the spherical micelles' release was faster than the release from the worm-like assemblies (Fig. \ref{encapsulation}). To quantify how the phase-transition affects the release, we continued measuring the release after the solutions' pH was adjusted from 6.5 to 4 or vice-versa, and the samples were placed in suitable fresh buffers. As controls, we used samples kept under the same initial pH and placed them into fresh buffers. It was fascinating to see the change in release rate due to the pH-induced shape transition as worm-like micelles that were transformed into spherical micelles (solid blue lines in Fig. \ref{encapsulation}a,b) started to release the dyes faster, showing a similar rate as the non-altered micelle control (dashed red lines in Fig. \ref{encapsulation}a,b). Simultaneously, a slower release was observed for spherical micelles, which were transformed into worm-like micelles (solid red lines and dashed blue lines in Fig. \ref{encapsulation}a,b). The changes in release rate due to the pH-induced change in mesophases were observed for both 2$\times$12 and 4$\times$7 IDPAs, with the effect being more significant for the latter. To ensure that the release rates were genuinely affected by the change in the assemblies' shape, we prepared a non-responsive amphiphile with two dodecane alkyl chains by replacing the hydrophilic IDP with a PEG chain of similar molecular weight. The PEG-based amphiphile (PEG-2$\times$12) self-assembled into micelles with a diameter of hydration of around 10 nm at both high and low pH (Supplementary Figure S5), demonstrating that its assembly is not responsive to pH at the tested range. We were encouraged to see that the control amphiphile release rates were not affected by the pH-jump after 3 hours. These results indicate that it is indeed the change in the shape of the IDPA assemblies due to altered interaction between the hydrophilic peptide domains, which affects the release rate.    

\subsection{Discussion}
We demonstrated the IDPA could serve as a valuable platform having stimuli-responsive self-assembly. Importantly, these nanoscopic assemblies are well-ordered, although originating from disordered peptides. When modeling the IDPA interactions and phase-transitions presented here, it is essential to understand whether the dominant interactions are non-specific, e.g., based on the total charge-density, or specific, e.g., conformational-based interactions. A generic model for a spherical-to-cylindrical micelle phase-transition will be agnostic to the charged peptide domain's details, aside from the charge distribution \cite{Lerche1987}. Such a model, detailed for brevity in the supplementary information, describes the free-energy balance between the energy associated with the hydrophilic region's charge-density and the energetic price paid for inducing a curvature that deviates from the intrinsic one. Essentially, lowering the pH results in a lower net charge that allows the hydrophilic peptide to compress. Assuming that the intrinsic curvature favors cylindrical morphology, this can facilitate the decrease in bending energy with a denser hydrophilic region. 

However, the mutated variants' experimental finding suggests that we must also consider the sequence's details and charges distribution along the peptide chain. Therefore, a sequence-dependent interaction term must be included in the free-energy calculation, describing the interaction between the amino-acids along two parallel and offset peptides \cite{Beck2010,Kornreich2015,Kornreich2016} (Supplementary Information). 

The experimental results on the \textit{IDPA2} variant show that indeed such a small change significantly affects the inter-micelle interactions, and as a result, the aggregation phase was dramatically changed. Therefore, the minute alteration between the original sequence and its variants suggests sequence-specific interactions. 
Moreover, the entire IDPA architecture, including the hydrophobic domain, also plays a role in the pH-response and the macroscopic aggregated mesophase (i.e., at low pH). For the 4$\times$7 hydrophobic domain, there is a clear structural difference between the \textit{IDPA2} and the other variants at high salt concentration. For the 2$\times$12 hydrophobic domain, while the condensed phase structural correlation weakens for the IDPA2 variants, they are salt-independent between 0.15-1 M. Indeed, as suggested by our theoretical modeling, the \textit{IDPA3} condensed phase (at pH=3) is somewhat similar to the original variant (see supplementary materials).

\section{Conclusions}

We investigated the conjugation of disordered polypeptide domains with hydrocarbon dendrimers into IDPA. We find that the interactions between the peptides lead to tunable self-assembled nanostructures. The IDPA hydrophilic domains are weakly interacting and disordered by nature, generally associated with transient behavior. Nonetheless, the IDPA self-assembly is remarkably forming ordered nanoparticles.

Moreover, the IDPA system shows pH response and structural phase-transition. It can be useful as a trigger for cargo release, similar to already demonstrated amphiphilic polymer-based nanoparticle systems with drug release enzymatic response. \cite{daniel2016dual}

We further demonstrated that minute alteration in the disordered peptide sequence and dendrimer architecture could significantly impact the nanoscopic and macroscopic length-scales. This platform also enables in-situ modifications of the IDPs and the study of the mutual interaction between IDPs. Last, as we showed here, modification of the IDP domain, and its interaction with the surrounding, plays a critical role in the IDPA's self-assembly and structural transformation from external cues. We expect that this platform can be furthered explored for targeted drug-delivery where tailored biological signals will induce phase-transition and expedite release.   

\section{Materials and methods}

\subsection{Synthesis and purification}
All peptides were synthesized at the Blavatnik Center for Drug Discovery (BCDD) at Tel Aviv University using automated Fmoc solid-phase peptide synthesis using The Liberty Blue$^{TM}$ automated microwave peptide synthesizer (CEM, Matthews, NC, USA). After the coupling of the last amino acid, either 3,5-bis(allyloxy)benzoic acid or 3,5-bis(prop-2-yn-1-yloxy)benzoic acid \cite{Harnoy2018} were coupled to the N-termius of the peptide. The capped peptides were cleaved from the resin usign standard conditions (95\% trifluoroacetic acid (TFA) (v/v), 2.5\% H2O (v/v), and 2.5\% triisopropylsilane (v/v) for 3 hours).The cleaved di-allyl or di-propargyl were purified by Waters AutoPurification system$^{TM}$ (MS directed LC) and were further reacted in thiol-ene or thiol-yne reactions with 1-dodecanethiol or 1-heptanethiol, respectively, as described below to yield IDPAs 2 $\times$ 12 and 4 $\times$ 7, respectively.

\ul{IDPA1-diallyl}: Exact Mass 2164.88. Detected Mass: ES mode +: 723.28 [(M+3H)/3] and 1084.21 [(M+2H)/2].

\ul{IDPA1-2$\times$12}: 54 mg (0.025 mmol) of peptide 2 was dissolved with 350$\mu$l of phosphate buffer (pH 7.4, 100mM) using gentle hitting, then 1 ml of DMF was added. 5.1 mg of DMPA (0.02 mmol) was separately solubilized with 100 $\mu$l of DMF and added to the peptide solution flowed by addition of 239.5 $\mu$l 1-dodecanethiol (0.997 mmol). The solution was purged with N$_2$ for 15 minutes and then stirred under UV light for 2 hours. Next, the crud mixture was placed in dialysis membrane with MWCO of 3,000 Da and dialyzed against DI water for 12 hr. The solution was lyophilized and final purification was performed using preparative-scale reversed-phase HPLC (Waters AutoPurification system). The product was confirmed by LC/MS. ACN was removed by rotary evaporation and the solution was further lyophilized yielding a white solid product 31.26 mg (48\% yield). Exact mass 2569.23. Detected mass: ES mode +: 857.98 [(M+3H)/3] and 1286.40 [(M+2H)/2].

\ul{IDPA1-dipropargyl}: Exact Mass 2160.85. Detected mass: ES mode +: 721.91 [(M+3H)/3] and 1082.23 [(M+2H)/2]. 

\ul{IDPA1-4$\times$7}: 58 mg (0.027 mmol) of peptide 1 were dissolved with 350$\mu$l of phosphate buffer (pH 7.4, 100mM) using gentle hitting, followed by addition o f 1ml of DMF. 5.48 mg of 2,2-Dimethoxy-2-phenylacetophenone (DMPA, 0.021 mmol) was separately solubilized with 100 $\mu$l of DMF and added to the peptide solution flowed by addition of 338 $\mu$l 1-heptanethiol (2.1 mmol). The solution was purged with N$_2$ for 15 minutes and then stirred under UV light for 2 hours. Next, the crud mixture was placed in dialysis membrane with MWCO of 3,000 Da and dialyzed against DI water for 12 hr. The solution was lyophilized and the product was purified using reparative-scale reversed-phase HPLC (Waters AutoPurification system). The product was confirmed by LC/MS. ACN was removed by rotary evaporation and further lyophilized yielding a white solid product 19.2 mg (26\% yield). Exact Mass 2689.24. Detected Mass: ES mode +: 898.17 [(M+3H)/3] and 1346.72 [(M+2H)/2].

\ul{IDPA2-diallyl}: Exact Mass 2164.88. Detected Mass: ES mode +: 723.26 [(M+3H)/3] and 1084.19 [(M+2H)/2]. 

\ul{IDPA2-2$\times$12}: Was synthesized similarly to IDPA1. 51 mg (0.024 mmol) of IDP2- were mixed with 4.8 mg of DMPA and 226 $\mu$l of 1-dodecanethiol and reacted and purified as was described for the synthesis of IDPA1 to yield 30.7 mg (50\% yield). MS analysis:  Exact Mass: 2569.23 Detected Mass: ES mode +: 858.17 [(M+3H)/3] and 1286.78 [(M+2H)/2].

\ul{IDPA2-dipropargyl}: Exact Mass 2160.85. Detected Mass: ES mode +: 721.89 [(M+3H)/3] and 1082.14 [(M+2H)/2]. 

\ul{IDPA2-4$\times$7}: Was synthesized and purified similarly to IDPA1. 54 mg (0.025 mmol) of IDP2-dipropargyl were mixed with 5.1 mg of DMPA and 315 $\mu$l of 1-heptanethiol and reacted and purified as was described for the synthesis of IDPA1 to yield 26.8 mg (40\% yield). MS analysis: Exact Mass: 2689.24 Detected Mass: ES mode +: 1346.78 [(M+2H)/2] and 898.22 [(M+3H)/3].

\ul{IDPA3-dipropargyl}: Exact Mass 2080.81. Detected Mass: ES mode +: 695.06 [(M+3H)/3] and 1041.95 [(M+2H)/2]. 

\ul{IDPA3-4$\times$7}: Was synthesized and purified similarly to IDPA1. 14 mg (0.0067 mmol) of IDP3-dipropargyl were mixed with 1.36 mg of DMPA and 84 $\mu$l of 1-heptanethiol, and reacted and purified as was described for the synthesis of IDPA1 to yield 9 mg (51\%). MS analysis:  Exact Mass: 2609.20 Detected Mass: ES mode +: 1306.83 [(M+2H)/2] and 871.47 [(M+3H)/3].

\ul{PEG-Dendron (2$\times$12) synthesis}: Dendron: 600 mg (2.5 mmol) of 3,5-(allyloxy) benzoic acid,\cite{Harnoy2018} 3.11 g of 1-dodecanethiol (15 mmol) and  38.4 mg of 2,2-Dimethoxy-2-
phenylacetophenone (DMPA; 0.15 mmol) were dissolved in 800$\mu$l of DMF. The solution was purged with N2 for 15 minutes and then stirred under UV light for 2 hours. Next, crud mixture was loaded on silica column, thiol excess was washed with 10:90 ethyl acetate and hexane (v/v) and 2$\times$12 dendron compound was eluted with 30:70 ethyl acetate and hexane (v/v), The fractions that contained the product were unified,evaporated and dried under high vacuum obtaining 1.5 g yellowish oily compound (92\% yield). 1H NMR (400 MHz, Chloroform-d) $\delta$ 7.23 (d, \emph{J} = 2.2 Hz, 2H), 6.69 (t, \emph{J} = 2.3 Hz, 1H), 4.10 (t, \emph{J} = 6.1 Hz, 4H), 2.70 (t, \emph{J} = 7.1 Hz, 4H), 2.52 (t, \emph{J} = 7.0 Hz, 4H), 2.07 (p, \emph{J} = 6.5 Hz, 4H), 1.59 (p, \emph{J} = 7.3 Hz, 4H),1.26-1.38 (m, 37H), 0.88 (t, \emph{J} = 6.8 Hz, 6H).
PEG-Dendron (2$\times$12): 70 mg of 2kDa PEG-amine \cite{Harnoy2014} were dissolved in 100 $\mu$l of DCM and 67.3 mg of 2$\times$12 dendron, and 40 mg of (2-(1H-benzotriazol-1-yl)-1,1,3,3-tetramethyluronium hexafluorophosphate HBTU were dissolved in DCM:DMF 1:1 (1mL) followed by addition of DIPEA 60 $\mu$l  were added to the PEG-amine solution and allowed to stir for 3 hour at room temperature. The crude mixture was loaded on a MeOH based LH20 SEC column. The fractions that contained the product were unified and the MeOH was evaporated in vacuum to obtain 76 mg (82\%) of the PEG based amphiphile.  
1H NMR (400 MHz, Chloroform-d) $\delta$ 6.91 (d, \emph{J} = 2.2 Hz, 2H), 6.69 (t, \emph{J} = 5.8 Hz, 1H), 6.57 (t, \emph{J} = 2.2 Hz, 1H), 4.08 (t, \emph{J} = 6.1 Hz, 4H), 3.83-3.45 (PEG backbone), 3.38 (s, 3H), 2.77 (t, \emph{J} = 6.5 Hz, 2H), 2.70-2.63 (m, 6H), 2.52 (t, \emph{J} = 7.4 Hz, 4H), 2.05 (p, \emph{J} = 6.6 Hz, H), 1.88 (p, \emph{J} = 6.6 Hz, 2H), 1.62 – 1.52 (m, 4H), 1.45 – 1.26 (m, 40H), 0.88 (t, \emph{J} = 6.7 Hz, 6H).

\subsection{SAXS and cryo-TEM sample preparation}
The IDPA or peptide powder was first fluidized in purified water (Milli-Q) at twice the desired concentration. The solution was then titrated with NaOH to a pH where the solution became more homogeneous (preferably a pH where the IDPAs are soluble in water). Titration was monitored using a pH probe (Sentek P13 pH Electrode). Following titration, 50 $\mu l$ of the solution was combined with 50 $\mu l$ of 2X buffer of choice to achieve a pH in the vicinity to the desired one. The 2X buffers Acetic Acid (pH 3-4.5), MES pH (5-6.5), and MOPS (pH 7-7.5) were prepared at 200 mM to achieve final buffer molarity of 100 mM after mixing with IDPA or peptide solution 1:1 (vol:vol).

\subsection{SAXS}
For solubilizing conditions (above the transition pH, generally above pH 6), samples were measured at three synchrotron facilities: Beamline B21, Diamond Light Source, UK, beamline 12.3.1, SIBYLS, Advanced Light Source, Berkeley, USA and beamline SWING, SOLEIL synchrotron facility, Paris, France. 

For phase-separating samples that display sediment (below the transition pH, generally pH 3-5.5), measurements were performed using an in-house X-ray scattering system, with a Genix3D (Xenocs) low divergence Cu $\rm{K}_\alpha$ radiation source (wavelength of $\lambda=1.54$~\AA) with a Pilatus 300K (Dectris) detector and scatterless slits setup\cite{li2008scatterless}, as well as beamline I22 at Diamond Light Source. Here, samples were measured inside 1.5 mm quartz capillaries (Hilgenberg).

\subsubsection{Peptide's SAXS analysis}
The unconjugated peptide by itself was measured using SAXS at different pH levels to test the effect on its ensemble-averaged structure. The peptide at each pH was measured at four different concentrations to extrapolate to the non-interacting "zero-concentration" peptides scatterings. From the low momentum transfer ($q$) regime of the extrapolated zero-concentration scattering curves, we extracted the radius of gyration ($R_g$) and the forward scattering, $I(0)$, using the Guinier analysis (Supplementary Figure S6). When examining the values of $R_g$ from high to low pH, it seems to remain a constant until pH $\sim$5.5 and gradually increase below it. However, this increase in size can be explained by a simultaneous increase in the effective mass of the peptides (increase in forwarding scattering) due to a decrease in inter-peptide repulsion near the isoelectric point (pI).

\subsection{Cryo-TEM}
Cryo-TEM specimen preparation was performed by applying a 6 $\mu l$ drop of the studied solution to a perforated carbon film supported on a 200-mesh TEM copper grid, thinning (blotting), and removing excess solution. Next, the sample was vitrified in liquid ethane at its freezing point (-183 \DC). The procedure was carried out at a controlled temperature (25 \DC) and water saturation. The vitrified specimens were stored under liquid nitrogen (-196 \DC) until examination. The samples were then examined using a Tecnai T12 G2 (FEI, The Netherlands) TEM operated at an accelerating voltage of 120kV, keeping specimen temperature below -170 \DC~ during transfer and observation. Images were digitally recorded on a Gatan Ultrascan 1000 cooled CCD camera using the Gatan Digital Micrograph software package. Images were recorded using methodologies we developed \cite{Danino2012} under low-electron-dose conditions to minimize electron beam radiation damage.

\subsection{Turbidity}
All measurements were recorded on a TECAN Infinite M200Pro device. The amphiphiles were treated and prepared in the same manner as previously described to achieve a final concentration of 5 mg ml-1. 100 $\mu l$ of each solution was loaded onto a 96 wells plate. The absorbance at 600 nm was scanned for each well.

\subsection {CD}
Circular dichroism (CD) measurements were performed using a commercial CD spectrometer (Applied Photophysics Chirascan). Both IDPAs, 2$\times$12 and 4$\times$7, and the unconjugated IDP, were placed in a glass cuvette with a 10 mm path length. The IDPAs and peptide were mixed with a phosphate buffer to achieve a concentration of 0.05 and 0.1 mg ml-1, respectively. Measurements were performed using phosphate buffer since the buffers used for the X-ray scattering experiments (mainly MOPS and MES) have high absorbance in the relevant CD wavelengths. The 190-260 nm wavelength range was probed in 1 nm steps, with 0.5 secs at each point. Three measurements were performed for each and averaged.

\subsection{CMC}
The amphiphile was dissolved in the diluent (15ml MOPS buffer solution (pH 7.4), 7.5$\mu$L of Nile red stock solution (2.5 mM in Ethanol) were added and mixed to give a final concentration of 1.25$\mu$M) to give the final concentration of 400$\mu$M IDPA and then sonicated for 5 minutes. This solution was repeatedly diluted by a factor of 1.5 with diluent. 100$\mu$L of each solution were loaded onto a 96 wells plate. The fluorescence emission intensity was scanned for each well (550 nm Emission intensity scan: 580-800 nm) using TECAN Infinite M200Pro plate reader. Maximum emission intensity was plotted vs. concentration in order to determine the CMC. This procedure was repeated thrice. 

\subsection{Cargo release experiment}

Release experiments of hydrophobic dye (butyl-coumarin) were performed using a dialysis tube (Mini GeBAflex-Kit, 8 kDa MWCO), volume range 10-250$\mu$L. A volume of 80$\mu$L of the micelles and worm-like micelle solutions were placed in the dialysis tubes, and each tube was immersed into 8 ml buffer at pH 6.5 or 4 with 0.5 mg/ml of BSA, respectively.

\begin{acknowledgement}

We thank Diamond Light Source for time on Beamlines B21 (SM24693) and I22 (SM21971), Advanced Light Source for time on Beamline SIBYLS 12.3.1 (SB-00941), and the SOLEIL synchrotron facility for time on Beamline SWING (20170798). The work was supported by the Israel Science Foundation (grant numbers 1454/20, 1553/18, 3292/19, 1117/16), Deutsche Forschungsgemeinschaft through SFB 958, NWU-TAU, the LMU-TAU collaboration initiatives, and DFG project GR1030/14-1. We thank Dr. Elvira Haimov and Dr. Boris Redko from the Blavatnik Center for Drug Discovery for their help with the synthesis and purification of the IDPAs.  
We also acknowledge fruitful discussions with Nathan Gianneschi, Joachim R{\"a}dler, Ram Avinery, and technical support of Mingming Zhang. 
\end{acknowledgement}

\begin{suppinfo}
The Supporting Information is available free of charge at XXX.
\end{suppinfo}

\bibliography{bibliography}
\end{document}